\documentclass[onecolumn]{aastex7}

\begin{document}

\title{High-inclination Centaur or Captured Halley-Type Comet? \\
Observational and Dynamical Characterization \\
of C/2024 L1 (PANSTARRS)}

\author[orcid=0000-0003-1008-7499, sname='T. Kareta']{Theodore Kareta}
\affiliation{Department of Astrophysics and Planetary Science, Villanova University, Villanova, PA, USA}
\email[show]{theodore.kareta@villanova.edu}  

\author[0000-0003-2152-6987]{John W. Noonan}
\affiliation{Department of Physics, Edmund C. Leach Science Center, Auburn University, Auburn, AL, USA}
\email{noonan@auburn.edu}

\author[0000-0003-1383-1578]{Benjamin N.L. Sharkey}
\affiliation{Department of Astronomy, University of Maryland, College Park, MD, USA}
\email{sharkey@umd.edu}

\begin{abstract}

The high-inclination (`high-\textit{i}') Centaurs of the middle and outer Solar System are of uncertain origin. While many have been captured directly from the Oort Cloud, making them less-evolved analogues to the Halley-Type Comets, this process alone cannot produce the number of observed objects. The most discussed alternate origin is direct excitation of trans-Neptunian Objects into high-\textit{i} orbits by the action of some unfound massive planet beyond Neptune. It is thus critical to find ways to discern the origins of these objects to provide new constraints on the various unfound planet hypotheses. This work assesses the only known active high-\textit{i} Centaur: C/2024 L1 (PANSTARRS). Our dynamical investigations show that L1 has not decreased its semimajor axis in at least the past 1810 years, unlike all known TNO-derived active Centaurs at lower-\textit{i} whose activity initiates after orbital changes. We also imaged L1 on five nights in the year before its April 2025 perihelion and found its brightness and optical colors to be stable and that its tail grew steadily. While this is like many Halley-type comets, few active low-\textit{i} Centaurs have such stable activity. We thus propose that L1 is a captured Oort Cloud comet analogous to lower perihelion Halley-types and is thus a highly attractive target for large aperture follow up. The number and active fraction of these high-\textit{i} objects should help us discern their origins and survey telescopes like Roman and Rubin could help us figure out where in the Outer Solar System the mystery lies.
\end{abstract}

\keywords{Comets, Centaurs, Oort Cloud Comets}

\section{Introduction} \label{sec:int}

\begin{figure*}[ht!]
\plotone{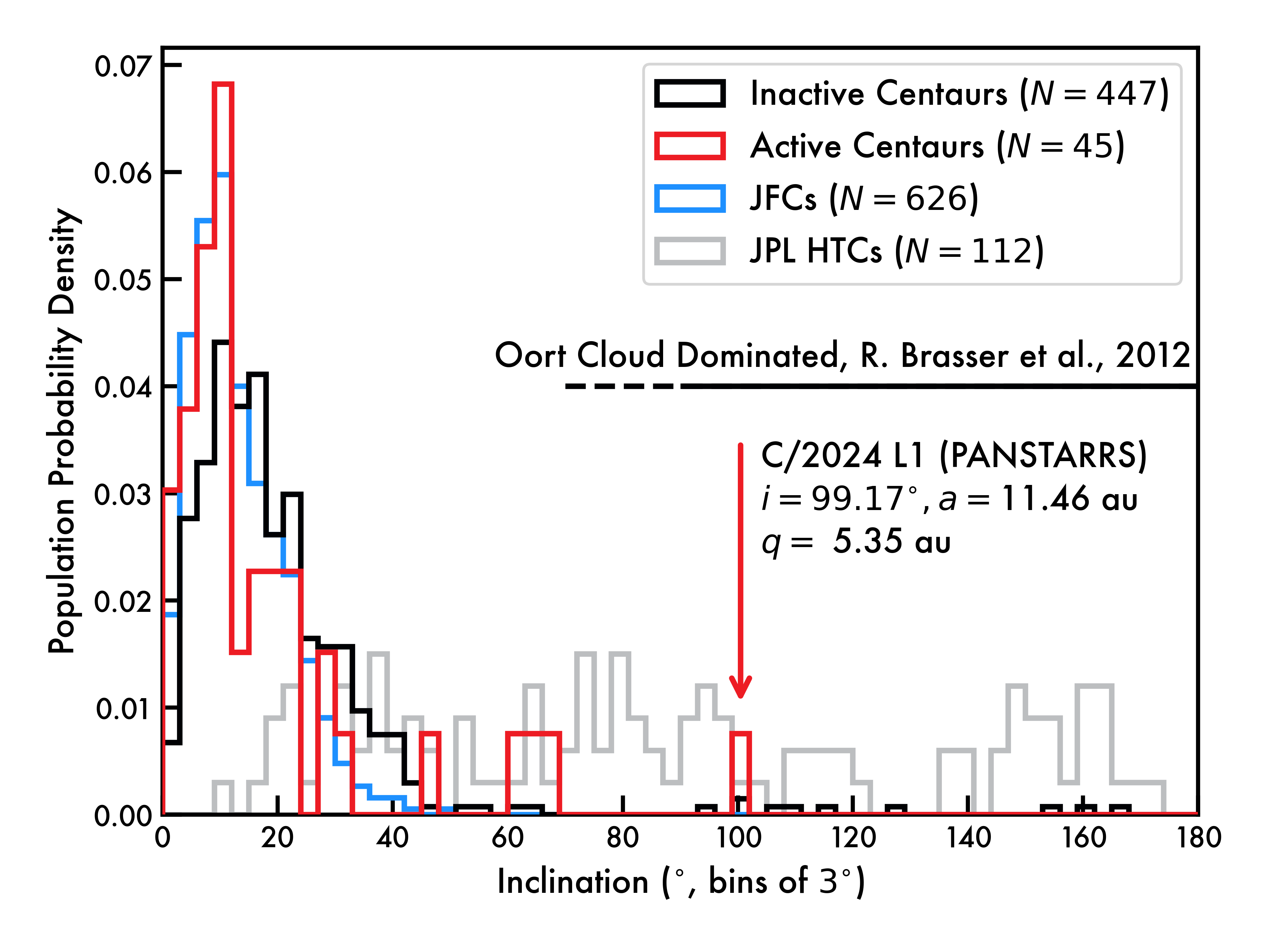}
\caption{The observed (not intrinsic) inclination distributions of several dynamical classes of comets and Centaurs are shown: inactive (black) and active (orange) Centaurs ($5.5 < a_{au} < 30.1$, $q_{au} > 5.2$ following \citet{2009AJ....137.4296J}), the JFCs in blue ($q_{au}<5.2$, $P_{years}<20$), and JPL's HTCs ($20<P_{years}<200$) in gray. C/2024 L1 (PANSTARRS) is the orange bump marked with an orange arrow near $100^{\circ}$. The data, downloaded from JPL's Small Bodies Database on 2026 October 31st, are shown as population-size and bin-width normalized histograms. Lastly, the estimates from \citet{2012MNRAS.420.3396B} of objects for whose inclinations are high enough to potentially need an Oort Cloud origin are shown as a black solid ($i>90^{\circ}$) and then dashed ($70^{\circ}<i<90^{\circ}$) line.}
\label{fig:i_dist}
\end{figure*}

The history of the Solar System is partially recorded in the current orbits of its small bodies. From the migration of the planets to how they were built and from what, discovering how the comets, asteroids, and trans-Neptunian Objects (TNOs) arrived at their present trajectories has been a powerful diagnostic. It is thus surprising that at least one population of small bodies remains challenging to identify the origin of: the high-inclination (hereafter just high-\textit{i}) Centaurs. While the low-\textit{i} Centaurs are on Giant Planet crossing orbits sourced directly from the TNOs (see, e.g., \citealt{1997Icar..127...13L, 2007Icar..190..224D, 2008ApJ...687..714V, 2015SSRv..197..191D, 2022arXiv220600010K}), the high-\textit{i} objects require a source or sources more widely distributed in inclination \citep{2012MNRAS.420.3396B,2019AJ....158...43K}. While repeated scattering off the Giant Planets and the recoil effect of cometary activity can both act to slowly raise the inclinations of comets and Centaurs, there are very few TNOs and JFCs with inclinations above $45^{\circ}$. The primary proposed pathways to produce the high-\textit{i} Centaurs in significant numbers is through direct capture from the Oort Cloud \citep{2012MNRAS.420.3396B,2019AJ....158...43K,2020MNRAS.497L..46M}.  However, careful debiasing of telescopic surveys has shown that the Oort Cloud alone cannot be the sole source of the high-\textit{i} Centaurs \citep{2019AJ....158..132N}) -- there are more objects known than predicted. An alternative process which may supply more high-\textit{i} objects is through TNOs getting excited to high-\textit{i} orbits by some yet-discovered mechanism, of which an unfound Giant Planet (\citealt{2019AJ....158..132N, 2023AJ....166..118L}, see also \citealt{2015Icar..258...37G, 2016ApJ...833L...3B}) somewhere beyond Neptune on an inclined orbit, is the most commonly discussed scenario. Such a planet has not been detected, but investigations and searches are ongoing. (Perhaps there are more inclined dwarf planets than anticipated? See, e.g., \citealt{2026arXiv261038693H}.)  Capture from the Oort Cloud would make these objects analogous to Halley-Type Comets (HTCs) with higher perihelia, while a TNO source should produce objects similar in composition and thermal state to those Centaurs in similar orbits at lower inclination. The inclination distribution of these populations of small bodies are shown in Figure \ref{fig:i_dist} for reference. In either case, the population is drawn from more distant reservoirs and subsequently preserved on these bizarre orbits. The properties and abundance of the high-\textit{i} Centaurs are thus a direct constraint on what processes might drive objects to high-\textit{i} in the Outer Solar System, including but not limited to the existence and orbit of an unfound planet, if observable differences can be found between objects sourced from either group in terms of their physical or orbital properties.

Unless stated otherwise in this paper, we adopt the following definitions and divisions between the dynamical classes of comets and icy Outer Solar System objects. Centaurs, regardless of activity state, are objects with semimajor axes $5.5<a<30.1$ and perihelia $q > 5.2$ following \citet{2009AJ....137.4296J}. JFCs are those comets with $P<20$ years, $2<T_J<3$, and $q<5.2$ au. HTCs have $20<P<200$ and $T_J<2$. There is thus an overlap in parameter space between Centaurs on high-inclination orbits and HTCs with sufficiently high perihelia, which we discuss throughout this paper.

There should be physical differences between high-\textit{i} Centaurs sourced from the Oort Cloud and the TNOs not just due to their different heliocentric distances ($a/au>10^3$ vs. $30<a/au<50$) resulting in different temperatures ($T/K<10$ vs. $30<T/K<60$) and subsequent thermal evolution and volatile loss patterns, but also because the comets that were implanted in both regions likely formed in distinct parts of the disk (see, e.g., \citealt{2015SSRv..197..191D}). However, the moderate to large geocentric distances of these objects when combined with the lack of spectral features on most comet and Centaur surfaces has made developing an observational way to assess the TNO vs. Oort Cloud balance challenging regardless of the nature of the differences between them. (This also contributes to significant  observational biases in which objects are even discovered -- fainter objects, and those on higher-inclination orbits, are both far less likely to be found.) Comet nuclei are generally red and featureless throughout the VIS/NIR wavelength range regardless of dynamical class \citep{2024come.book..361K}, though there are likely some weak trends within classes related to perihelion distance. The lack of any clustering or grouping of comet and active Centaur spectra, contrasted against the spectral diversity of the TNOs discussed previously, suggests that evolutionary processes might render the surfaces of these objects undiagnostic in terms of discerning their origin. JWST observations of the surfaces of two inactive high-inclination TNOs \citep{2025NatAs...9..230P}, 2012 DR30 and 2013 LU28, showed their surfaces to be Bowl-type / H$_2$O-rich like the few comet nuclei studied by JWST thus far (including objects both from the Oort Cloud and TNOs, see \citealt{2025MNRAS.541L...8S, 2026PSJ.....7..137S}). Given that Bowl-type / H$_2$O-rich surfaces seem to be able to come from both the Oort Cloud and the TNOs, and independently that comets sourced from both populations seem to show a broad overlap in reflective properties among the dozens studied, it seems possible that their surfaces alone might not encode enough information to unambiguously sort them into groups according to their origins. Even the current orbits of individual objects in the high-\textit{i} population are not obviously distinct enough to sort them into groups. While an object with a perihelion well inside the terrestrial planet region, high inclination, and a semimajor axis among the Giant Planets would be quickly and unambiguously classified as a Halley-Type comet, an object with the same orbit but with a perihelion above Jupiter's orbit is now plausibly described as a Centaur or Halley-type depending on the scientific question to be answered. (In other words, some authors might not even separate the two populations.) How, then, can we discern where these objects came from and better understand what might lie unfound beyond Neptune?

In this paper, we present a comprehensive observations-and-dynamics study of the brightest and by most definitions only active high-\textit{i} Centaur, C/2024 L1 (PANSTARRS), to assess the nature and cause of its cometary activity beyond the water ice line and to potentially infer which of these populations it might originate in as well as the differences between those populations. It is the only active Centaur with $i>90^{\circ}$. Cometary activity on ``L1" was discovered in PANSTARRS observations from 2025 June 1 ($R_H\sim5.62$ au and $\Delta\sim4.92$ au) where the object was seen to have a small coma less than an arcsecond wide. Active objects like ``L1", whose orbital elements are shown in Table \ref{tab:orb} and whose inclination is specifically marked in Figure \ref{fig:i_dist}, can have their properties, compositions, and internal structures ascertained through far more means than inactive objects -- by studying how they brighten and dim with heliocentric distance, how long their activity persists, and what kinds of solids and gases being emitted from their surfaces. Comets from different parts of the solar system show distinct chemistry and behavior and thus our goal is to investigate whether these differences can help us sort out the origins of individual high-\textit{i} objects. There are still relatively few comets of any type which have been studied in great detail at these distances so the nature of L1's activity is of interest regardless of whether it is more like an HTC or a low-\textit{i} Centaur.

%\subsection{Distinguishing Sources of Active Objects}
%The primary goal of this paper is to assess whether high-\textit{i} Centaurs which show cometary activity can be distinguished. Centaurs and HTCs both show cometary activity in the Giant Planet region where these high-\textit{i} objects are found, and thus activity is plausible to expect on these objects regardless of their source. Given that cometary activity allows for a different way to investigate the compositions and structures of these objects, perhaps we can use the atmospheres and tails of any high-\textit{i} Centaurs to probe what populations they originate in.

%There is only one active object which meets the \citet{2009AJ....137.4296J} Centaur definition ($5.5<a/{au}<30.1$, $q/au > 5.2$) while having an inclination high enough to pass the \citet{2012MNRAS.420.3396B} $i>90^{\circ}$ criteria: C/2024 L1 (PANSTARRS), whose orbital elements are shown in Table \ref{tab:orb} and which is the primary focus of this paper. The inclination distributions of the inactive Centaurs, the active Centaurs, the JFCs, and the HTCs are shown in Figure \ref{fig:i_dist} with C/2024 L1 marked for context. Of the 65 active Centaurs which meet the $q$ and $a$ criteria, only five ($\sim7.5\%$) have inclinations over $45^{\circ}$. C/2024 L1 is the only one which has $i>90^{\circ}$ in the whole sample. 

\begin{deluxetable*}{cccccc}[ht!]  \label{tab:orb}
\tablecaption{Orbital Elements for C/2024 L1 (PANSTARRS) taken from JPL orbital solution 12.}
\tablecolumns{6}
\tablewidth{0pt}
\tablehead{
\colhead{$a$ (au)} &
\colhead{$e$} & 
\colhead{$i$ ($^{\circ}$)} &
\colhead{$\Omega$ ($^{\circ}$)} &
\colhead{$\lambda$ ($^{\circ}$)} & 
\colhead{$t_p$ (date)}}
\startdata
11.46 & 0.533 & 99.17 & 45.13 & 171.83 & 2025 April 23.54 \\
\enddata
\end{deluxetable*}

%Even if separating the high-i HTCs from the high-i active Centaurs is a creating a false dichotomy, there isn't a comparable HTC with a comparably high inclination and perihelion distance. (If one queries JPL's Small Body Database (``SBDB") for comets with $20 <P_{years}<200$ (the JPL HTC definition) with the above cut on perihelion distance, one actually retrieves the same five objects\footnote{A cut based on $T_J$, the Tisserand parameter with respect to Jupiter, is not necessarily better. The lowest inclination of these five objects is the unusually bright C/2012 Q1 (Kowalski) with $a=26.13$ au, $q = 9.48$ au, and $i=45.18^{\circ}$ which is classified as a JFC based on its relatively mundane $T_J=2.634$. Considering that its MOID with respect to Jupiter is about Jupiter's semimajor axis, the object's dynamical similarities to traditional JFCs are few.} with $i>45^{\circ}$ -- the high-perihelion HTCs and the high-inclination Centaurs are only distinguishable in some contexts.) Every object with a comparable $q$ has too low an $i$ and vice-versa. In either case, the object is dynamically distinct from other active Centaurs and HTCs and easily meets the $i > 90^{\circ}$ threshold from \citet{2012MNRAS.420.3396B} to be sourced directly from the Oort Cloud.

The structure of this paper is as follows. In Section \ref{sec:orb}, we present our dynamical integration of L1's orbit over the past three millennia to understand its stability and when its activity might have initiated. In Section \ref{sec:obs}, we present multi-filter photometry and imaging of this comet covering the year prior to its April 2025 perihelion just exterior to Jupiter to characterize its modern activity state and dust coma. In Section \ref{sec:disc}, we synthesize the results of our dynamical and observational investigations into L1 to compare to other active Centaurs and Halley-Types and thus constrain what it can tell us about the structure of the far Outer Solar System.

\section{Recent Orbital Evolution} \label{sec:orb}
The origin, onset, and duration of cometary activity in the Outer Solar System is tightly correlated with how their orbits have changed -- at least for the active Centaurs \citep{2018P&SS..158....6F, 2024ApJ...960L...8L}. As comets age in a stable orbit, the volatile ices they have near to their surfaces are slowly lost to sublimation and a lag layer of dust and debris builds up which inhibits the sublimation of the remaining ices. Every time a Centaur is scattered into a warmer orbit (its semimajor axis decreases), temperatures are raised throughout the subsurface and new pockets of volatiles become ready to sublimate -- it seems that all of the low-\textit{i} active objects have had such orbital changes within the past few hundred years \citep{2024ApJ...960L...8L}. 

\begin{figure*}[ht!]
\plotone{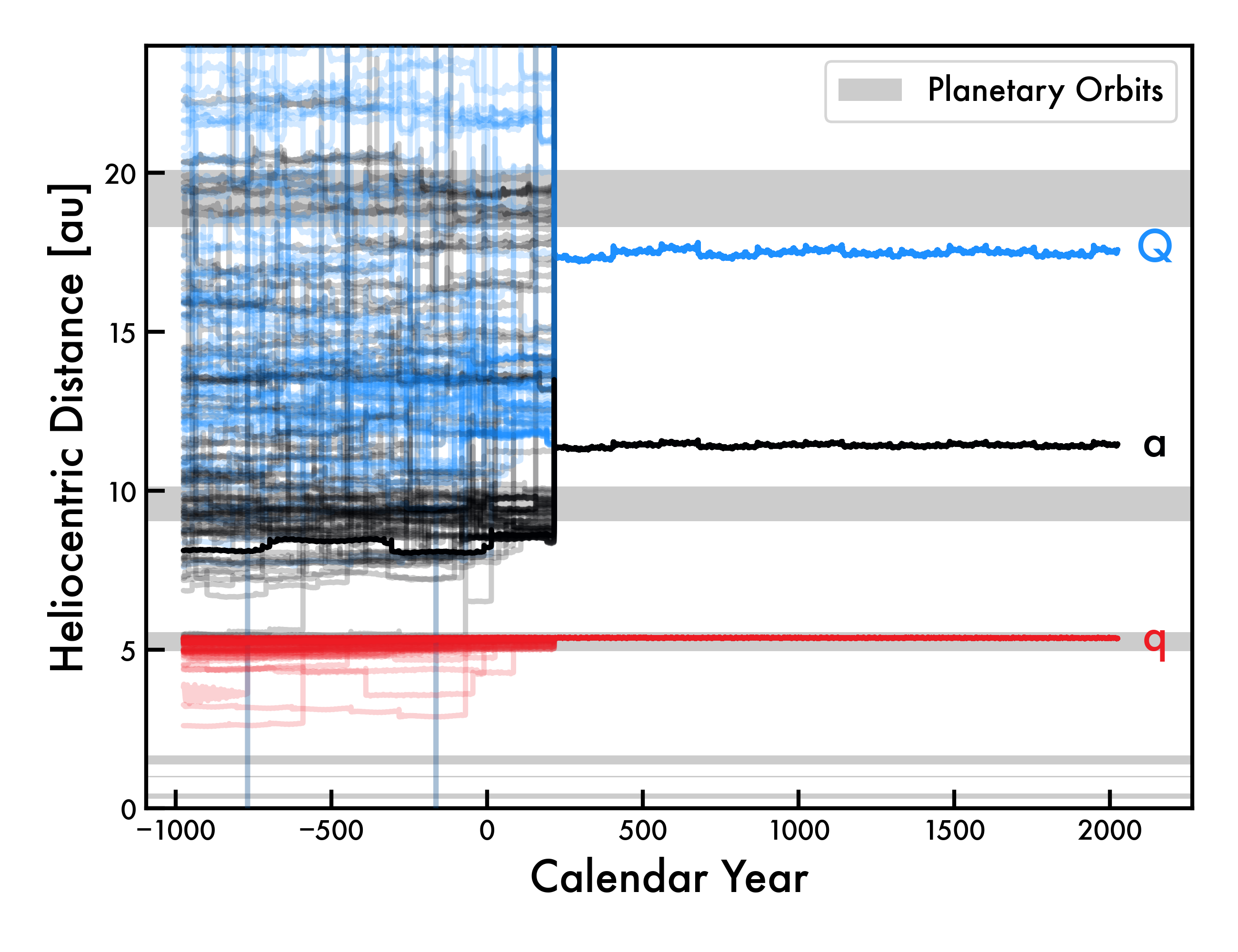}
\caption{A backward integration of C/2024 L1 (PANSTARRS) is shown as described in the text, with the perihelia of the nominal orbit and one hundred clones shown in red, their semimajor axes in black, and aphelia in blue.}
\label{fig:orb}
\end{figure*}

To investigate the recent history of C/2024 L1 (PANSTARRS), we input the nominal orbit from JPL (Solution 12, dated 12 August 2025) and one hundred `clones' sampled from the uncertainty matrix of the orbital solution into \textit{Rebound} \citep{2012A&A...537A.128R}, the widely-used open source N-body code. We utilized the IAS15 integrator \citep{2015MNRAS.446.1424R} which handles close encounters properly with a dynamic timestep as close encounters are expected to dominate the relatively rapid orbital evolution of the Centaurs. We integrated the nominal orbit and clones back three thousand years from the solution epoch with the position and orbital elements of each particle written out every $0.02$ years, or a little bit less than weekly. The evolution of the object's perihelion, semimajor axis, and aphelion is shown in Figure \ref{fig:orb}.

Compared to active Centaurs which have all had semimajor axis decreases in the past several hundred years, the high-inclination orbit of C/2024 L1 (PANSTARRS) means that it has close planetary encounters relatively rarely. It has a perihelion relatively close to Jupiter but is furthest from the plane of the Solar System when it is at a Saturn-like heliocentric distance, showing that Jupiter is still the dominant cause of its ongoing orbital evolution. While the perihelion distance of L1 has remained quite stable within the $5.34-5.38$ au range for the past $\sim1810$ years, the small changes that can be seen in its semimajor axis and aphelion distance are driven by weaker and more distant periodic encounters with Jupiter, Saturn, and Uranus in order of decreasing importance. Considering the differences in oribtal period of L1 and Jupiter, the two objects might be expected to have a close approach in the ecliptic every $\approx1900$ years (every $\sim49$ L1 orbits or every $\sim160$ Jupiter orbits), so the stability timeframe derived statistically from Rebound is likely to be real and not just a product of the current orbital uncertainty.

The trajectories of the orbital clones of L1 begin to diverge significantly prior to a close approach with Jupiter in 215 CE at $0.045$ AU or slightly less (determined with a finer $0.005$-year timestep) which raised the semimajor axis of most of L1's clones from a \textit{lower} value around Saturn's semimajor axis to the modern value of $11.46$ au. These lower-$a$ orbits also had lower eccentricities with near-Jupiter perihelia similar to the modern day, indicating that the short periods of warmth near perihelion are shorter than they had been prior and that the rest of the orbit is now somewhat colder as well. A very small fraction ($4/100$) had larger semimajor axes prior to 215 CE with a wider range of eccentricities. All orbital changes subsequent to that encounter are small by comparison. Either of these scenarios is in stark contrast to the low-inclination active Centaurs \citep{2018P&SS..158....6F, 2024ApJ...960L...8L} which all show drops in their semimajor axes within the several hundreds years prior to the discovery of their activity. Even post-encounter, there have been multiple stretches of a hundred years or more in which L1's semimajor axis was smaller than it is today. In other words, model of Centaur activity initiation based on orbital changes which make new volatile pockets accessible to sublimation would not necessarily predict activity on this object at present. This is evidence that the origin and duration of C/2024 L1's activity is distinct from centaurs in similarly shaped orbits but at lower inclinations, potentially driven by differences in evolutionary state and/or source region. We return to this topic in Section \ref{sec:disc}.

We also specifically investigated the spread in orbital inclination among the clones as the backward integration proceeded. In general, 99 of the 100 total clones retained an inclination higher than $90^{\circ}$ till at least 1000 BCE (``before the common era"). The only clone which previously had $i<90^{\circ}$ did so well after the other orbital parameters $a$ and $e$ had already begun to de-cohere. Back-integrations of Solar System objects are only reliable until they begin to spread out rapidly in parameter space (see, e.g., \citealt{2020MNRAS.497L..46M}), and so we did not investigate L1's trajectory significantly earlier than this. The timescale for von Zeipel-Kozai-Lidov oscillations for an object like L1 orbiting the Sun perturbed by Jupiter is $\approx3800$ years (see, e.g., \citealt{2013degn.book.....M}), and yet even within our 3000 year window we see no obvious oscillation in eccentricity at this or other periods. We thus conclude that C/2024 L1 has had an orbit nearly-perpendicular to plane of the Solar System for at least several thousand years and likely longer.

\section{Multi-Epoch Imaging} \label{sec:obs}
Starting in June 2024 and leading up to L1's perihelion in 2025 April, we imaged C/2024 L1 (PANSTARRS) on multiple occasions with the Lowell Discovery Telescope (hereafter `LDT') with a variety of filters summarized in Table \ref{tab:obs}. L1 was a dim and extended target (peaking near an integrated magnitude $m_V\approx21$ around opposition after being discovered approximately one magnitude fainter), so observations were only attempted on nights with good or excellent conditions. We utilized the Large Monolithic Imager (hereafter `LMI', \citealt{2014SPIE.9147E..2NB}) in 3$\times$3 binning mode (pixel scale $0.36"$) for all dates included in Table \ref{tab:obs} except for the 2025 April run which utilized 2$\times$2 binning ($0.24"$/pixel). We debiased and flatfielded our images using standard techniques identically on all dates.  Astrometric registration of all images and photometric calibration of images taken through the Sloan $g'$, $r'$, $i'$, $z'$ filters and the broadband $VR$ filter was done through PhotometryPipeline \citep{2017A&C....18...47M}. We only utilized field stars with visible colors similar to those of the Sun for calibration of those broadband filters. We also utilized a HB $CN$ filter \citep{2000Icar..147..180F} in our 2024 June observations for which we also observed a bright standard star (HD 120086) for which that filter had been calibrated. However, no object was detected at the expected position of the comet in individual frames nor in a stack of them and thus they are excluded from much further analysis. We tentatively estimate an upper limit on the production rate of CN gas of $Q_{CN} < 5 \times10^{25}$ mols/s (e.g., it is at least a little less active than comet Hale-Bopp was at similar distances, which is unsurprising), but stress that these numbers need to be taken with a grain of salt. CN emission is only beginning at Jupiter-like heliocentric distances and is thus typically detected in only the brightest and most active comets (see, e.g., \citealt{1997Sci...275.1918W}), so our non-detection is unsurprising and more difficult to interpret than if the comet was well within the water sublimation zone.

\begin{deluxetable*}{c|c|c|c|c|c|c}[ht!]  \label{tab:obs}
\tablecaption{Log of Lowell Discovery Telescope Imaging Observations.}
\tablecolumns{7}
\tablewidth{0pt}
\tablehead{
\colhead{Date} &
\colhead{$R_H$ (au)} & 
\colhead{$\Delta$ (au)} & 
\colhead{$\alpha$ ($^{\circ}$)} & 
\colhead{Filter)} &
\colhead{$N_{exp}\times \tau_{exp}$ (s)} &
\colhead{Seeing ("), Conditions}}
\startdata
2024 June 11.21 - 11.29 & 5.61 & 4.96 & 8.44 & $g'$, $r'$, $i'$, $z'$ & $10\times120$s & 0.8, Clear \\
- & - & - & - & $CN$ & $5\times300$s & -\\
2024 Aug. 13.15 - 13.21 & 5.52 & 5.61 & 10.41 & $VR$ & $80\times45$s & 0.9, Variable Clouds\\
2024 Sep. 6.12 - 6.15 & 5.49 & 5.91 & 9.22 & $VR$ & $25\times60$s & 1.3-1.4, Clear\\
2025 Mar. 25.45-25.47 & 5.35 & 4.68 & 8.42 & $g'$, $r'$, $i'$, $z'$ & 90s color sequence & 1.5, Clear \\
2025 Apr. 25.40-25.43 & 5.35 & 4.38 & 3.14 & $g'$, $r'$, $i'$, $z'$ & 2 90s color sequences & 0.85, Clear \\
\enddata
\tablecaption{Log of Lowell Discovery Telescope Imaging Observations.}
\end{deluxetable*}

\subsection{Photometry and Colors}

\begin{figure*}[ht!]
\plotone{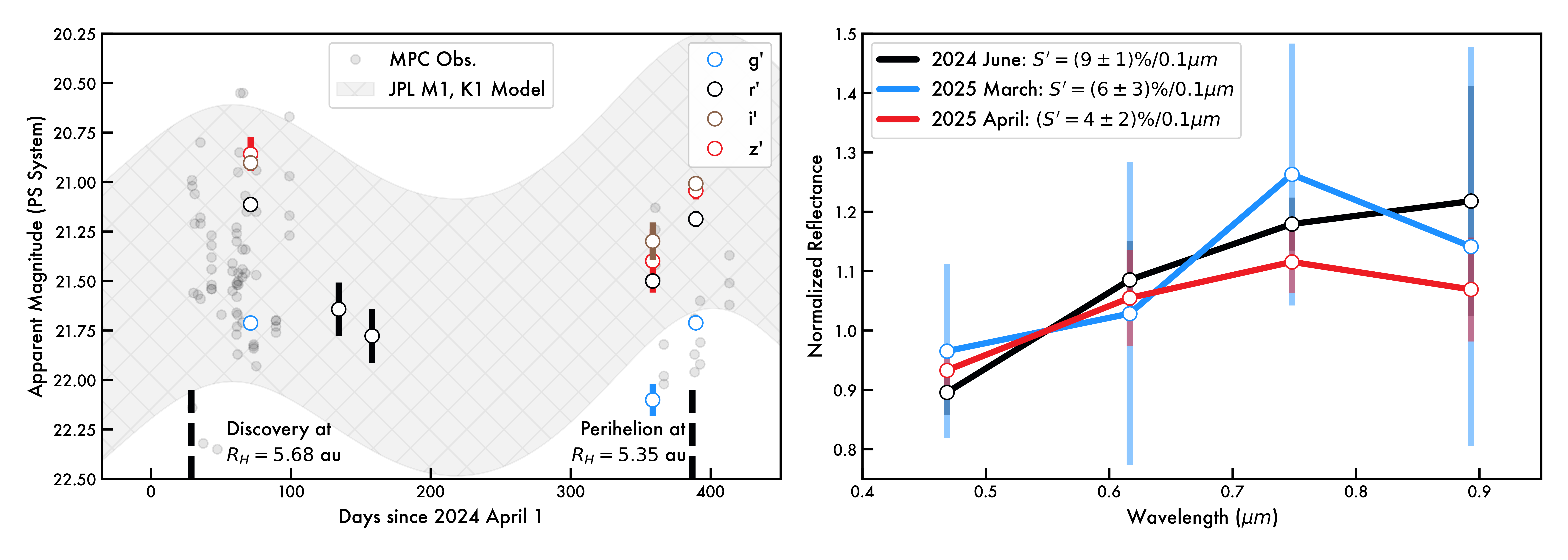}
\caption{Left: The overall brightness evolution of C/2024 L1 (PANSTARRS) is shown, with new $g'r'i'z'$ photometry from the Lowell Discovery Telescope shown as colored unfilled circles and photometry accepted at the Minor Planet Center shown as high-opacity smaller gray circles. The predicted magnitude from JPL Horizons is shown as a gray area whose height indicates uncertainties behind the photometry. The discovery of L1 in 2024 April and its later perihelion in 2025 April are noted with dashed vertical lines at the bottom. Right: Our new $g'r'i'z'$ photometry converted to reflectance values and normalized at $0.55\mu{m}$.}
\label{fig:phot}
\end{figure*}

As a result of the object's substantial heliocentric distance and relative dimness, the comet was never very widely extended on-sky even though its dust tail grew in length over time. In a stack of all of the Sloan $r'$ frames from 2024 June, the FWHM of the comet was $1.1-1.2$ arcseconds compared to $0.75-0.85$ arcseconds for nearby field stars. The size of the coma compared to the size of the seeing disk, even when considering our conditions in 2024 June, 2024 August, and 2025 April were significantly better than what is typically achieved at the LDT, thus precluding detailed investigation of the morphology of L1's inner coma. We thus extracted photometry of L1's coma with a $3.6-$arcsecond radius ($3\times$FWHM) circular aperture on all nights and in all filters. The results of our five nights of multi-filter photometry are shown in Figure \ref{fig:phot} plotted alongside photometry from the Minor Planet Center and the JPL Horizons predicted magnitude of the object. 

Our photometry is clearly consistent with other photometry taken around the same time, but is often on the brighter end of the MPC observation envelope likely due to our extraction aperture being wider than what is typically used for point sources. One clear trend shown in both the MPC observations and our own is that the comet was brighter during its discovery epoch (2024 April to July, $R_H\sim5.6$ au and $\Delta\sim5$ au) than it was at perihelion a year later ($R_H\sim5.35$ au and $\Delta\sim4.38$ au) when it was $\sim0.35$ au closer to the Sun. The JPL Horizons standard comet model of the object's brightness has considerable uncertainty in its total magnitude ($M1_V=14.3\pm0.7$, far larger than typical photometric errors, and no estimated error on its brightening slope, $k = 4.5$) in part because of this. We thus attempted to fit our own standard M1, K1 model to our own higher-SNR $r'$ data for comparison. We retrieved similarly uncertain fits with brighter absolute magnitudes and weaker dependence on heliocentric distance ($M1\sim15.2$, $k\sim4$), but we view these as similarly unreliable first-guesses. The comet simply would be expected to brighten as it moved towards perihelion by about a third of a magnitude, but dimmed very slightly over the same time period. All of this leads us to conclude that L1 was potentially discovered in a small sub-magnitude outburst or a state of slightly elevated activity and that its 2025 brightnesses (approximately $0.1$ magnitudes fainter) might be more indicative of its overall activity state. We searched for precovery images prior to its official discovery using the B612 Institute's \textit{ADAM} tool with no success, which may be soft evidence for the brightening starting around when the object was discovered and thus a key reason it was discovered then.

\begin{deluxetable*}{c|c|c|c|c|c}[ht!]  \label{tab:phot}
\tablecaption{Photometry of C/2024 L1 (PANSTARRS) between 2024 and 2025.}
\tablecolumns{6}
\tablewidth{0pt}
\tablehead{
\colhead{Date} &
\colhead{$m_r$} & 
\colhead{$m_g$} &
\colhead{$m_i$} &
\colhead{$m_z$} & 
\colhead{$Af\rho$ $(m_r)$ (cm)}}
\startdata
2024 June 11.21 - 11.29 & $21.11\pm0.03$ & $21.72\pm0.02$ & $20.90\pm0.02$ & $20.85\pm0.09$ & $33\pm1$ \\
2024 Aug. 13.15 - 13.21 & $21.64\pm0.13$ & - & - & - & $22\pm3$\\
2024 Sep. 6.12 - 6.15 & $21.77\pm0.14$ & - & - & - & $21\pm3$\\
2025 Mar. 25.45-25.47 & $21.50\pm0.03$ & $22.10\pm0.08$ & $21.30\pm0.09$ & $21.40\pm0.16$ & $20\pm1$ \\
2025 Apr. 25.40-25.43 & $21.19\pm0.04$ & $21.71\pm0.03$ & $21.01\pm0.03$ & $21.04\pm0.04$ & $25\pm1$\\
\enddata
\end{deluxetable*}

The right side of Figure \ref{fig:phot} shows the evolution of how this comet's reflectivity has changed during three nights in which multi-color photometry were obtained. As can be seen in the plot, the normalized reflectivities derived from our 2024 June, 2025 March, and 2025 April observations all overlap with each other quite closely. We can instead consider their overall spectral slopes using the $S'$ formalism \citep{1990Icar...86...69L}, we retrieve $S'=9\pm1$ in 2024 June, $S'=6\pm3$ in 2025 March, and $S'=4\pm2$ in 2025 April all in units of $\%/0.1\mu{m}$. This indicates that between the time of a potential increase in activity around its discovery and when it reached perihelion a year later, this comet became slightly more spectrally neutral at the approximately $2-\sigma$ level, below the common $3-\sigma$ threshold used in our field. If the $z'$ photometry is excluded simply for the reason that it is the least certain, the difference in reflectivities over the year timeframe are even smaller: we retrieve $S'=9\pm1$ in 2024 June, $S'=10\pm3$ in 2025 March, and $S'=6\pm1$ in 2025 April in $\%/0.1\mu{m}$. In other words, if L1 was indeed in a state of elevated activity as the various sources of photometry might suggest, the difference between that state and its later activity at perihelion did not meaningfully change the way its dust coma reflected light and thus no major changes in its dust population are likely to have occurred.

\subsubsection{Nucleus Size Estimation}
While our observations did not observe the nucleus directly, we can still use our own observations and those submitted to the Minor Planet Center to constrain the size of it. As mentioned elsewhere in the paper, the comet dimmed slightly from 2024-2025, and thus we take the dimmest observations accepted by the MPC around its perihelion to be the most constraining. These observations are around $m_V\approx22$, which after correcting for distance and phase angle allow us to estimate an upper limit on the brightness of the nucleus $H_V>15.4$. This corresponds to an upper limit on the object's diameter of $D <  5.5$ km, which is larger than both the average JFC and LPC \citep{2017AJ....154...53B}. Given that CN production is active for some comets and inactive for others at this distance as CN production is often tied to the sublimation of water ice, our loose CN production rate upper limit is less useful and less reliable in determining L1's size.

\subsection{Morphology}
\begin{figure*}[ht!]
\plotone{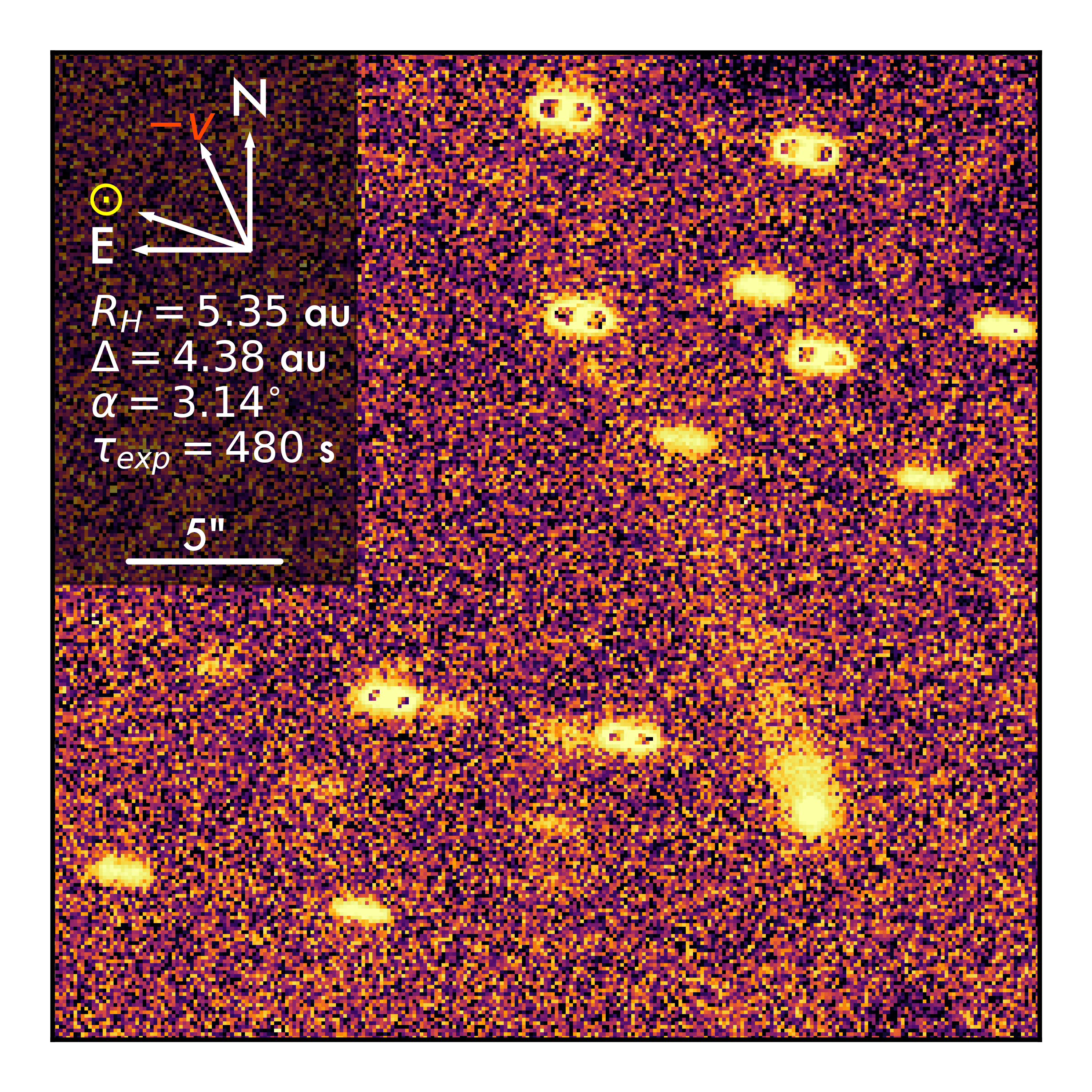}
\caption{A deep stack of C/2024 L1 (PANSTARRS) in the Sloan $r'$ filter is shown taken on 2025 April 25 with the Lowell Discovery Telescope two days after its perihelion. The comet's tail can be seen to be at least $\sim30"$ long, but due to the small phase angle and large heliocentric distance we expect the apparent tail to be significantly foreshortened (e.g., most of it is 'into the page'). The comet's tail grew steadily in length as it approached its near-Jupiter perihelion.}
\label{fig:stack}
\end{figure*}

While even deep stacks of our 2024 June data on L1 when it was intrinsically brightest barely show a tail at all, this changed significantly over the next year as it approached perihelion. By 2024 September 6, L1's tail had grown to be at least $\sim15$ arcseconds long in a deep $VR$ stack and by 2025 April 25, L1's tail it was at least $\sim32$ arcseconds long in a deep stack in the Sloan $r'$ filter. The latter of these two stacks is shown as Figure \ref{fig:stack}.

Given that the tail is longest in the latest images taken at the lowest phase angles where it should be foreshortened the most (e.g., the tail should primarily point `into the page' and thus appear shorter when projected onto the sky) and is challenging to discern at all in the images taken near its discovery when the comet was at more than double the phase angle, it seems like all or most of this tail growth was driven by the comet's dust output as it approached perihelion and not from changes in viewing geometry. Given that the comet's near-nucleus brightness (a proxy for the object's current dust production rate, see again Figure \ref{fig:phot}) did not change or slightly dimmed over the year prior to its 2025 perihelion, it seems possible that the object's growing tail length simply reflects that the comet has been active longer. In other words, it is essentially a proxy for how long the comet has been producing dust this apparition and thus L1 only really started its activity around when it was discovered in early 2024. This aligns with the lack of precovery detections of the object. We return to the topic of when the comet activated and how might reflect its origins later on.

The tail is somewhat less visible in longer wavelength filters in the April 2025 data, which is likely due to a combination of lower SNR in those filters due to the brightness of the Sun and sensitivity of the detector as well as the decreased reflectivity of any smaller grains at longer wavelengths should they be present. At these large distances, all but the smallest ejected dust should largely move in the comet's anti-velocity direction  (see, e.g., \citealt{1968ApJ...154..327F, 2014acm..conf..565V}) due to the slow action of radiation pressure. Syndynes (see again \citealt{1968ApJ...154..327F}) plotted on the comets position largely overlap regardless of the synthetic particles' susceptibility to radiation pressure. While we cannot very effectively characterize its dust size distribution, we can rule out a population dominated by ultra-fine grains which would be still meaningfully effected by solar radiation pressure at these distances.

\section{Discussion} \label{sec:disc}
Few comets as far away as L1 have been monitored over a long enough timescale to see their activity evolve in response to changing solar insolation, and the vast majority of those observational campaigns have been focused for scientific and capability reasons on brighter and larger comets (see, e.g., \citealt{1989Icar...77..223B, 2001Icar..150...94B, 2008A&A...485..599T, 2022ApJ...933L..44K}). L1 is on the fainter end, and thus the smaller end, of the kinds of comets that have been studied so far out and thus might be expected to be more `typical' of the broader comet population's properties than the larger objects like 95P/Chiron or Comet Bernardinelli-Bernstein. In addition to simply characterizing the way that this distant object on an unusual orbit behaves and evolved, the two primary goals of this Discussion section are to attempt to ascertain where C/2024 L1 (PANSTARRS) came from in the Outer Solar System before arriving at its present orbit and to begin to develop a plan and set of questions to be asked as large telescopes like the Vera Rubin Observatory begin to peer deeper into the enigmatic population of high-\textit{i} Centaurs of which L1 is the only known active member.

\subsection{Origin of C/2024 L1 (PANSTARRS)}
As mentioned in Section \ref{sec:int}, the one well-understood source of the high-\textit{i} Centaurs is through direct implantation from the Oort Cloud, analogous to high-\textit{q} Halley-Type Comets \citep{2012MNRAS.420.3396B, 2019AJ....158...43K, 2020MNRAS.497L..46M}. However, the apparent size of the population of high-\textit{i} Centaurs is too large in number for the Oort Cloud to be the sole source \citep{2019AJ....158..132N}. While there are ways to produce high-\textit{i} Centaurs which are unlikely to show activity (e.g., they have lost their volatiles or never had many to begin with, \citealt{2020AJ....160..144G, 2018P&SS..158....6F}), excitation of TNOs onto high-\textit{i} orbits by some undiscovered planet on an inclined orbit could do so \citep{2019AJ....158..132N, 2023AJ....166..118L} should it exist. How activity is initiated, sustained, and evolves on comets sourced from the Oort Cloud and the trans-Neptunian objects is distinct, and thus the nature of L1's activity and how it started is a constraint on the structure of the Outer Solar System.

In \citet{2024ApJ...960L...8L}, those authors found that all currently active Centaurs and high-perihelion JFCs had significant decreases in their semimajor axes over the past millennium, and the vast majority have had their orbital changes within the past 250 years or less. Activity seems to start relatively rapidly (within months) for many objects as new pockets of ice become warm enough to sublimate, though longer lags of years to decades for some objects may be possible depending on how one interprets the biased historical record of such faint objects. While it is not clear how long activity persists once started by this mechanism, we can infer from these constraints that activity can last several hundred years but most probably not more than a thousand after a change in $a$ as these objects slowly lose their surface volatiles to space. This kind of analysis would not predict activity on C/2024 L1, and thus its ongoing activity is different in terms of its apparent cause and duration compared to the activity of the lower-inclination Centaurs. 

In contrast, it is not unreasonable to expect that a HTC sourced from the Oort Cloud even thousands of years ago would still be active, as even the namesake 1P/Halley is detectably active at these distances \citep{1991plas.rept..173M} despite having at least 2200 years of strong activity in the Inner Solar System \citep{1999ccc..book.....K}. Given that Halley's warmer perihelion should result in a \textit{faster} loss of more volatile ices which are likely to be powering L1's modern activity, Halley's modern activity beyond the water ice line is good evidence that these objects can retain hypervolatile-based activity for at least several thousand years and likely longer in colder orbits like that of L1. Halley's nucleus is relatively large (D$\sim$11 km) for a comet, and clearly still capable of activity on long timescales, so L1 does not necessarily need to be similar in size to Centaurs like 29P/Schwassmann-Wachmann 1 or 174P/Echeclus to sustain activity over thousands of years; it may be smaller than Halley. L1's long-term persistent activity given its lack of recent orbital changes is only hard to explain if it were derived from the TNOs, but not if it were derived from the Oort Cloud like the Halley-Type comets closer to the Sun.

Impacts between comets and other bodies (often smaller ones) are sometimes discussed as a way to reinitiate activity on formerly active but still ice-bearing objects. Due to the high inclination of L1 we cannot use the classical \"{O}pik or Wetherill collisional models \citep{1951PRIA...54..165O, Wetherill1967}, which assume uniform nodal precession and low-inclination target orbits. To investigate the likelihood of collisional activation, we implemented the minimum orbit intersection distance Monte Carlo collision probability method from \citet{Rickman2014}, which allows us to calculate the impact probability between the highly inclined L1 and the Centaur and Jupiter Trojan populations to verify our assumption that the orbit of C/2024 L1 was indeed extremely unfavorable to small impacts which might expose new patches or pockets of ice. Given the small sizes of both the target and expected impactors, gravitational focusing was not expected to be a significant factor, and was omitted \citep[e.g.][]{1985Icar...63..290W}. After calibrating the model using main belt asteroids and comparing to intrinsic collision probabilities from \citet{bottke_collisional_2015} we then calculated an order-of-magnitude collision rate for C/2024 L1. We modeled collision rates for the current orbit of C/2024 L1 with a sample of all Centaurs pulled from the Minor Planet Center (MPC), then debiased by $sin(i)\times a^2$ scaled by a total number of Centaurs r$>$0.5 km estimated at 214,000, roughly scaled from the estimated 21,400 Centaurs with r$>$10 km in \citet{2025AJ....169...73K}. We used a similar methodology to account for the Jupiter Trojan population, sampling 10,000 known Trojans from the MPC and scaling to 600,000 total Trojans with r$>$0.5 km for both L4/L5 populations, a crude approximation from the results of \citet{Jewitt2000} but a reasonable guess. Based on the derived absolute magnitude of $\sim$14.3 L1 likely has a radius of 4 km, assuming a cometary albedo of 0.05, but we round up to 5 km allowing us to maximize the collision rate. Even with this scaling, the collisional frequency is $\sim$3$\times$10$^{-12}$ yr$^{-1}$, dominated largely by the Trojans ($>98\%$), as L1 crosses the ecliptic at 5.4 au. If we were to crudely extend a size frequency power law of $r^{-2}$ down to the 10 meter size range for the Trojans, the collision rate would approach 10$^{-8}$ yr$^{-1}$; still far too infrequent to explain activity in the present day.  In other words, based on the recent evolution of L1's orbit \textit{and the fact that it is active at all}, this object appears distinct from the TNO-derived low-\textit{i} Centaurs closer to the plane of the Solar System.

The overall stability of L1's activity state is also weak evidence towards its physical properties being more similar to HTCs than low-\textit{i} Centaurs -- at least, to their overall population averages. L1's activity appears to have initiated sometime around or slightly before its discovery in early 2024 judging by the growth of its tail and initially small size on-sky. While L1 was slightly brighter at opposition one year prior to perihelion than it was at perihelion, its tail seems to have grown stably and consistently over this timeframe indicating a general stability (or, at least, continuity) of its dust production rate. The color of L1's dust was also generally stable over this timeframe (see Fig. \ref{fig:phot}), with variations at the $2-\sigma$ level if all of our $g'r'i'z'$ photometry is included and at even lower significance if the lowest-SNR $z'$ is discarded. The color of the dust was also largely typical for comets with a weak red slope. In other words, L1 appears to have released similar kinds of dust stably for at least a year continuously without any obvious outbursts, and certainly no large ones. While there \textit{are} Centaurs with near-Jupiter perihelia which also show stable activity within some critical $R_H$ (see, e.g., a preliminary report in \citealt{2024AGUFMP24B...02F}), these objects seem to be relatively uncommon among the broader population of active Centaurs. Many active Centaurs show highly stochastic patterns of activity with limited, if any, correlation with their heliocentric distance and thus temperature. L1's relatively stable activity without outbursts or changes in its dust population is more like the HTCs than it is the low-\textit{i} active Centaurs, but there is clearly some amount of overlap in the properties of the two populations.

An important caveat, however, is that there are significant observational biases in our knowledge of the Active Centaurs. Specifically, it might be possible that the larger objects, like 29P/Schwassmann-Wachmann 1 or 174P/Echeclus, are more likely to be outburst-dominated than smaller objects which are more challenging to study due to their faintness. In any case, L1's activity is qualitatively distinct from many of the best-studied active Centaurs if not most of them. A fuller census of active Centaur activity profiles, especially of fainter objects, will be possible with the Vera Rubin Observatory's LSST and will facilitate a more meaningful discussion of the activity state of these rarer objects in the coming years. If an active high-\textit{i} Centaur were to be found to have more low-\textit{i} Centaur like characteristics, that could be used to argue for the ongoing transport of TNO material to high-\textit{i} orbits -- and thus potentially for the existences of processes to do so -- but this current object does not meet that requirement.

Combining the dynamical and observational evidence, then, it seems most likely that C/2024 L1 (PANSTARRS) is essentially a Halley-Type Comet preserved at large heliocentric distance -- the exact kind of object that \citet{2012MNRAS.420.3396B} predicted would be in this population. Considering that the population otherwise has no unambiguous active members and that activity on HTCs can persist for millennia, perhaps it is one of the younger members of the population.

Monitoring L1 as it recedes from the Sun to see how its activity changes, where it ceases, and to monitor for any outbursts might add additional context to this discussion that might aid in sorting it into the Halley-type-like or Centaur-like (or other) categories. Observations at other wavelengths which might be able to tap into the aforementioned chemical differences between the two proposed sources of this object, especially with longer-wavelength facilities like JWST, could be critical in taking these fascinating rare objects and turning them fully into a way to constrain the existence or not of an undiscovered planet beyond Neptune. Given the relative faintness of the object, near-infrared photometry or visible spectroscopy might be possible from the largest facilities but space-based assets are likely critical to further characterize it.

\subsection{Where are the other active high-i objects?}
Why don't we see more active high-i Centaurs? Is that a constraint on the existence of an unfound object at high-\textit{i} or a population thereof, or is it an effect of small number statistics? Even if we assume that most of the objects sourced from the Oort Cloud behave largely like Halley does, maintaining stable activity relatively far out for a few thousand years, and the objects sourced from the trans-Neptunian belt behaved like the active Centaurs do, only becoming active after significant changes in their orbit, then the fact that we have only found one active object \textit{could be} unsurprising for a few reasons. 

First off, the high-\textit{i} objects simply do not encounter the planets very frequently due to the differences in their orbital planes. Even if every single one of them was excited into their current orbit by some unfound planet, we would only have a chance to see activity on those which have recently had a rare encounter which lowered their orbits. Second off, this argument (that one can tell the compositions of TNOs and Oort Cloud Objects apart) implicitly assumes that the TNOs which are excited onto Centaur orbits by Neptune and by an unfound planet are not from such different parts of the trans-Neptunian region as to have radically different composition. Estimating the active fraction of high-\textit{i} Centaurs as a function of where an unfound planet might be (e.g., how different compositions in the TNO belt might map onto activity durations and detectability) might help assess this. As an extreme example, a mostly-water-ice object from the Haumea \citep{2007Natur.446..294B, 2012AJ....143..146B} family might not show much activity at all in the Centaur region, and perhaps other populations of TNOs might show distinct patterns here that our results are sensitive to.

Lastly, perhaps Halley's larger size means that it can maintain activity for much longer than is typical among this population -- if the population were to be dominated by smaller objects (as is expected, \citealt{2017AJ....154...53B}), their activity might only last for a fraction of Halley's two millenia and thus the population could be both water-ice-rich and mostly dormant. That said, L1 is not as bright as Halley at the same distance -- and is thus likely smaller -- and hasn't had an orbital change in nearly two millennia. L1 spends about $\sim15\%$ of its orbit within the CO$_2$ snowline and has been directly observed to be active for at least $\sim3\%$ of its orbit so far. If L1 is typical of these Oort Cloud injected objects as we have proposed here, there could be somewhere between $6-30$ undiscovered objects with identical properties that we haven't detected yet because they are currently beyond where their activity would make them detectable. As described in Section 1.1, the known population of high-\textit{i} objects is between these numbers (about 15 or so\footnote{Specifically, if JPL's small body database is queried for objects with $5.2<a<30$, $q>5.2$, and $i>45^{\circ}$, it returns fourteen inactive objects and L1.}) and C/2024 L1 is still the only uncontestable member of it which is active. Given even a slightly larger sample size, the number of active high-\textit{i} objects could thus become a very useful lever arm in understanding. Whether the mystery lies at 45 au or 45,000 au, the lack of active high-\textit{i} objects should be a high priority to investigate with Rubin and other facilities like Roman in the coming years.

\section{Summary}
The high-inclination Centaurs of the middle and outer Solar System are of uncertain origin. While many are likely to be captured Oort Cloud Comets, analogous to Halley Type Comets with much larger perihelia \citep{2012MNRAS.420.3396B}, this process is unlikely to produce enough objects to explain the known population \citep{2019AJ....158..132N}. The primary alternative hypothesis for how to deliver them is through some alternative method to excite TNOs to high-\textit{i} orbits at large $R_H$, including but not limited to the action of an unfound massive object on an inclined orbit \citep{2019AJ....158..132N, 2023AJ....166..118L}. The size and properties of the population of active Centaurs is thus a constraint on the structure of the far Outer Solar System, or at least the processes currently operating within it.

In this paper, we present a dynamical and telescopic investigation of the one known high-\textit{i} Centaur which shows cometary activity, C/2024 L1 (PANSTARRS), to attempt to discern whether it came from the Oort Cloud like the Halley-Types or from the trans-Neptunian belt like the lower-\textit{i} Centaurs. Few comets of any dynamical class have been followed for a substantial fraction of their orbit at these large heliocentric distances, and many of those have been much larger and brighter comets, so the properties of L1 are of interest regardless of how it is classified.

Our dynamical investigation provides support both for L1's high-\textit{i} orbit being long-lived (e.g., not a result of von Zeipel-Kozai-Lidov oscillations) and for its semimajor axis and perihelion remaining similarly unchanged for nearly the past two millennia. All known active Centaurs at lower \textit{i} have moved into orbits with smaller semimajor axes within the past few hundred years, so the object's current activity makes it clearly distinct dynamically from active objects at lower \textit{i} which are sourced from the TNOs. Cometary activity on the lower-\textit{i} Centaurs seems to only persist for several hundred years after a move into a warmer orbit, while L1 may have been active in this orbit for two millenia or longer indicating significant physical differences between L1 and those objects at lower-\textit{i}.

The comet reached a maximum integrated brightness around $m\approx21$ and our five nights observations were focused on multi-filter imaging starting shortly after discovery and followed the target until its 2025 perihelion. The observational evidence also points towards the Halley-Types and away from the classical Centaurs. While HTCs frequently display stable cometary activity at these distances (e.g., mass loss correlated with heliocentric distance), Centaurs seem to do so much less frequently as opposed to showing stochastic and variable activity with less clear correlation with heliocentric distance. L1's activity evolved slowly and continuously without any detected outbursts, its tail grew steadily over the year prior to its perihelion, and the color of its dust comet did not change within uncertainties.

We conclude on the basis of the dynamical and observational evidence that C/2024 L1 (PANSTARRS) is much more likely to be a high-perihelion Halley-Type comet analog than a TNO-derived Centaur excited to higher inclination by some unknown mechanism. These kinds of analyses can be repeated with all high-\textit{i} Centaurs that the Vera Rubin Observatory's LSST finds, but finding more ways to break the observational ambiguities is clearly required if these active high-\textit{i} Centaurs are to be useful in the hunt for or case against an unfound planet or other inclination excitation processes beyond Neptune.

\begin{acknowledgments}
We thank Audrey Thirouin among others for useful conversations which improved this paper. We thank Aeralyn Flynn, Arya Menk, and Katie Breeland-Newcomb for assistance with the 2024 June observations as well as Lowell Discovery Telescope operators Ishara Nisely, Benjamin Shafransky, and Cecilia Siqueiros for their expertise and support throughout all five observing runs. Lastly, we thank two anonymous reviewers which significantly improved many aspects of this manuscript. This work was supported in part by internal Villanova University funds.

\end{acknowledgments}

\begin{contribution}
TK was responsible for data acquisition, reduction and analyses in addition to drafting and editing the paper. JWN and BNLS assisted in interpretation of the data and in preparation of the text of this manuscript.

\end{contribution}

\facilities{LDT(LMI)}

\software{\textit{Rebound} \citep{2012A&A...537A.128R, 2015MNRAS.446.1424R}, \textit{photometrypipeline}, \citep{2017A&C....18...47M}}

%\bibliography{sample7}{}

\begin{thebibliography}{}
\expandafter\ifx\csname natexlab\endcsname\relax\def\natexlab#1{#1}\fi
\providecommand{\url}[1]{\href{#1}{#1}}
\providecommand{\dodoi}[1]{doi:~\href{http://doi.org/#1}{\nolinkurl{#1}}}
\providecommand{\doeprint}[1]{\href{http://ascl.net/#1}{\nolinkurl{http://ascl.net/#1}}}
\providecommand{\doarXiv}[1]{\href{https://arxiv.org/abs/#1}{\nolinkurl{https://arxiv.org/abs/#1}}}

\bibitem[{{Batygin} \& {Brown}(2016)}]{2016ApJ...833L...3B}
{Batygin}, K., \& {Brown}, M.~E. 2016, \apjl, 833, L3, \dodoi{10.3847/2041-8205/833/1/L3}

\bibitem[{{Bauer} {et~al.}(2017){Bauer}, {Grav}, {Fern{\'a}ndez}, {Mainzer}, {Kramer}, {Masiero}, {Spahr}, {Nugent}, {Stevenson}, {Meech}, {Cutri}, {Lisse}, {Walker}, {Dailey}, {Rosser}, {Krings}, {Ruecker}, {Wright}, \& {NEOWISE Team}}]{2017AJ....154...53B}
{Bauer}, J.~M., {Grav}, T., {Fern{\'a}ndez}, Y.~R., {et~al.} 2017, \aj, 154, 53, \dodoi{10.3847/1538-3881/aa72df}

\bibitem[{{Bida} {et~al.}(2014){Bida}, {Dunham}, {Massey}, \& {Roe}}]{2014SPIE.9147E..2NB}
{Bida}, T.~A., {Dunham}, E.~W., {Massey}, P., \& {Roe}, H.~G. 2014, in Society of Photo-Optical Instrumentation Engineers (SPIE) Conference Series, Vol. 9147, Ground-based and Airborne Instrumentation for Astronomy V, ed. S.~K. {Ramsay}, I.~S. {McLean}, \& H.~{Takami}, 91472N, \dodoi{10.1117/12.2056872}

\bibitem[{Bottke {et~al.}(2015)Bottke, Brož, O’Brien, Campo~Bagatin, Morbidelli, \& Marchi}]{bottke_collisional_2015}
Bottke, W.~F., Brož, M., O’Brien, D.~P., {et~al.} 2015, in Asteroids {IV}, ed. P.~Michel, F.~E. DeMeo, \& W.~F. Bottke (University of Arizona Press), \dodoi{10.2458/azu_uapress_9780816532131-ch036}

\bibitem[{{Brasser} {et~al.}(2012){Brasser}, {Schwamb}, {Lykawka}, \& {Gomes}}]{2012MNRAS.420.3396B}
{Brasser}, R., {Schwamb}, M.~E., {Lykawka}, P.~S., \& {Gomes}, R.~S. 2012, \mnras, 420, 3396, \dodoi{10.1111/j.1365-2966.2011.20264.x}

\bibitem[{{Brown} {et~al.}(2007){Brown}, {Barkume}, {Ragozzine}, \& {Schaller}}]{2007Natur.446..294B}
{Brown}, M.~E., {Barkume}, K.~M., {Ragozzine}, D., \& {Schaller}, E.~L. 2007, \nat, 446, 294, \dodoi{10.1038/nature05619}

\bibitem[{{Brown} {et~al.}(2012){Brown}, {Schaller}, \& {Fraser}}]{2012AJ....143..146B}
{Brown}, M.~E., {Schaller}, E.~L., \& {Fraser}, W.~C. 2012, \aj, 143, 146, \dodoi{10.1088/0004-6256/143/6/146}

\bibitem[{{Bus} {et~al.}(2001){Bus}, {A'Hearn}, {Bowell}, \& {Stern}}]{2001Icar..150...94B}
{Bus}, S.~J., {A'Hearn}, M.~F., {Bowell}, E., \& {Stern}, S.~A. 2001, \icarus, 150, 94, \dodoi{10.1006/icar.2000.6555}

\bibitem[{{Bus} {et~al.}(1989){Bus}, {Bowell}, {Harris}, \& {Hewitt}}]{1989Icar...77..223B}
{Bus}, S.~J., {Bowell}, E., {Harris}, A.~W., \& {Hewitt}, A.~V. 1989, \icarus, 77, 223, \dodoi{10.1016/0019-1035(89)90087-0}

\bibitem[{{Di Sisto} \& {Brunini}(2007)}]{2007Icar..190..224D}
{Di Sisto}, R.~P., \& {Brunini}, A. 2007, \icarus, 190, 224, \dodoi{10.1016/j.icarus.2007.02.012}

\bibitem[{{Dones} {et~al.}(2015){Dones}, {Brasser}, {Kaib}, \& {Rickman}}]{2015SSRv..197..191D}
{Dones}, L., {Brasser}, R., {Kaib}, N., \& {Rickman}, H. 2015, \ssr, 197, 191, \dodoi{10.1007/s11214-015-0223-2}

\bibitem[{{Farnham} {et~al.}(2000){Farnham}, {Schleicher}, \& {A'Hearn}}]{2000Icar..147..180F}
{Farnham}, T.~L., {Schleicher}, D.~G., \& {A'Hearn}, M.~F. 2000, \icarus, 147, 180, \dodoi{10.1006/icar.2000.6420}

\bibitem[{{Fern{\'a}ndez} {et~al.}(2018){Fern{\'a}ndez}, {Helal}, \& {Gallardo}}]{2018P&SS..158....6F}
{Fern{\'a}ndez}, J.~A., {Helal}, M., \& {Gallardo}, T. 2018, \planss, 158, 6, \dodoi{10.1016/j.pss.2018.05.013}

\bibitem[{{Fernandez} {et~al.}(2024){Fernandez}, {Schambeau}, {Lisse}, {Bolin}, \& {Bauer}}]{2024AGUFMP24B...02F}
{Fernandez}, Y.~R., {Schambeau}, C.~A., {Lisse}, C.~M., {Bolin}, B., \& {Bauer}, J. 2024, in AGU Fall Meeting Abstracts, Vol. 2024, AGU Fall Meeting Abstracts, P24B--02

\bibitem[{{Finson} \& {Probstein}(1968)}]{1968ApJ...154..327F}
{Finson}, M.~L., \& {Probstein}, R.~F. 1968, \apj, 154, 327, \dodoi{10.1086/149761}

\bibitem[{{Gomes} {et~al.}(2015){Gomes}, {Soares}, \& {Brasser}}]{2015Icar..258...37G}
{Gomes}, R.~S., {Soares}, J.~S., \& {Brasser}, R. 2015, \icarus, 258, 37, \dodoi{10.1016/j.icarus.2015.06.020}

\bibitem[{{Greenstreet} {et~al.}(2020){Greenstreet}, {Gladman}, \& {Ngo}}]{2020AJ....160..144G}
{Greenstreet}, S., {Gladman}, B., \& {Ngo}, H. 2020, \aj, 160, 144, \dodoi{10.3847/1538-3881/aba2c9}

\bibitem[{{Holman} {et~al.}(2026){Holman}, {Juri{\'c}}, {Napier}, {Kurlander}, {Heinze}, {Moeyens}, {Schwamb}, {Bernardinelli}, {Sullivan}, {Bellm}, \& et~al.}]{2026arXiv261038693H}
{Holman}, M.~J., {Juri{\'c}}, M., {Napier}, K.~J., {et~al.} 2026, arXiv e-prints, arXiv:2610.38693.
\newblock \doarXiv{2610.38693}

\bibitem[{{Jewitt}(2009)}]{2009AJ....137.4296J}
{Jewitt}, D. 2009, \aj, 137, 4296, \dodoi{10.1088/0004-6256/137/5/4296}

\bibitem[{{Jewitt} {et~al.}(2000){Jewitt}, {Trujillo}, \& {Luu}}]{Jewitt2000}
{Jewitt}, D.~C., {Trujillo}, C.~A., \& {Luu}, J.~X. 2000, \aj, 120, 1140, \dodoi{10.1086/301453}

\bibitem[{{Kaib} \& {Volk}(2022)}]{2022arXiv220600010K}
{Kaib}, N.~A., \& {Volk}, K. 2022, arXiv e-prints, arXiv:2206.00010, \dodoi{10.48550/arXiv.2206.00010}

\bibitem[{{Kaib} {et~al.}(2019){Kaib}, {Pike}, {Lawler}, {Kovalik}, {Brown}, {Alexandersen}, {Bannister}, {Gladman}, \& {Petit}}]{2019AJ....158...43K}
{Kaib}, N.~A., {Pike}, R., {Lawler}, S., {et~al.} 2019, \aj, 158, 43, \dodoi{10.3847/1538-3881/ab2383}

\bibitem[{{Kelley} {et~al.}(2022){Kelley}, {Kokotanekova}, {Holt}, {Protopapa}, {Bodewits}, {Knight}, {Lister}, {Usher}, {Chatelain}, {Gomez}, {Greenstreet}, {Angel}, \& {Wooding}}]{2022ApJ...933L..44K}
{Kelley}, M. S.~P., {Kokotanekova}, R., {Holt}, C.~E., {et~al.} 2022, \apjl, 933, L44, \dodoi{10.3847/2041-8213/ac7bec}

\bibitem[{{Knight} {et~al.}(2024){Knight}, {Kokotanekova}, \& {Samarasinha}}]{2024come.book..361K}
{Knight}, M.~M., {Kokotanekova}, R., \& {Samarasinha}, N.~H. 2024, in Comets III, ed. K.~J. {Meech}, M.~R. {Combi}, D.~{Bockel{\'e}e-Morvan}, S.~N. {Raymond}, \& M.~E. {Zolensky}, 361--404, \dodoi{10.2458/azu_uapress_9780816553631-ch012}

\bibitem[{{Kronk}(1999)}]{1999ccc..book.....K}
{Kronk}, G.~W. 1999, {Cometography: A Catalog of Comets, Volume 1: Ancient-1799}

\bibitem[{{Kurlander} {et~al.}(2025){Kurlander}, {Holman}, {Bernardinelli}, {Juri{\'c}}, {Heinze}, \& {Payne}}]{2025AJ....169...73K}
{Kurlander}, J.~A., {Holman}, M.~J., {Bernardinelli}, P.~H., {et~al.} 2025, \aj, 169, 73, \dodoi{10.3847/1538-3881/ad9a58}

\bibitem[{{Levison} \& {Duncan}(1997)}]{1997Icar..127...13L}
{Levison}, H.~F., \& {Duncan}, M.~J. 1997, \icarus, 127, 13, \dodoi{10.1006/icar.1996.5637}

\bibitem[{{Lilly} {et~al.}(2024){Lilly}, {Jev{\v{c}}{\'a}k}, {Schambeau}, {Volk}, {Steckloff}, {Hsieh}, {Fernandez}, {Bauer}, {Weryk}, \& {Wainscoat}}]{2024ApJ...960L...8L}
{Lilly}, E., {Jev{\v{c}}{\'a}k}, P., {Schambeau}, C., {et~al.} 2024, \apjl, 960, L8, \dodoi{10.3847/2041-8213/ad1606}

\bibitem[{{Luu} \& {Jewitt}(1990)}]{1990Icar...86...69L}
{Luu}, J., \& {Jewitt}, D. 1990, \icarus, 86, 69, \dodoi{10.1016/0019-1035(90)90199-J}

\bibitem[{{Lykawka} \& {Ito}(2023)}]{2023AJ....166..118L}
{Lykawka}, P.~S., \& {Ito}, T. 2023, \aj, 166, 118, \dodoi{10.3847/1538-3881/aceaf0}

\bibitem[{{Meech}(1991)}]{1991plas.rept..173M}
{Meech}, K.~J. 1991, {Outburst of Comet Halley at 14.3 AU}, In NASA, Washington, Reports of Planetary Astronomy, 1991 p 173 (SEE N92-12792 03-89)

\bibitem[{{Merritt}(2013)}]{2013degn.book.....M}
{Merritt}, D. 2013, {Dynamics and Evolution of Galactic Nuclei}

\bibitem[{{Mommert}(2017)}]{2017A&C....18...47M}
{Mommert}, M. 2017, Astronomy and Computing, 18, 47, \dodoi{10.1016/j.ascom.2016.11.002}

\bibitem[{{Morbidelli} {et~al.}(2020){Morbidelli}, {Batygin}, {Brasser}, \& {Raymond}}]{2020MNRAS.497L..46M}
{Morbidelli}, A., {Batygin}, K., {Brasser}, R., \& {Raymond}, S.~N. 2020, \mnras, 497, L46, \dodoi{10.1093/mnrasl/slaa111}

\bibitem[{{Nesvorn{\'y}} {et~al.}(2019){Nesvorn{\'y}}, {Vokrouhlick{\'y}}, {Stern}, {Davidsson}, {Bannister}, {Volk}, {Chen}, {Gladman}, {Kavelaars}, {Petit}, {Gwyn}, \& {Alexandersen}}]{2019AJ....158..132N}
{Nesvorn{\'y}}, D., {Vokrouhlick{\'y}}, D., {Stern}, A.~S., {et~al.} 2019, \aj, 158, 132, \dodoi{10.3847/1538-3881/ab3651}

\bibitem[{{{\"O}pik}(1951)}]{1951PRIA...54..165O}
{{\"O}pik}, E.~J. 1951, Proceedings of the Royal Irish Academy, 54, 165

\bibitem[{{Pinilla-Alonso} {et~al.}(2025){Pinilla-Alonso}, {Brunetto}, {De Pr{\'a}}, {Holler}, {H{\'e}nault}, {Feliciano}, {Lorenzi}, {Pendleton}, {Cruikshank}, {M{\"u}ller}, {Stansberry}, {Emery}, {Schambeau}, {Licandro}, {Harvison}, {McClure}, {Guilbert-Lepoutre}, {Peixinho}, {Bannister}, \& {Wong}}]{2025NatAs...9..230P}
{Pinilla-Alonso}, N., {Brunetto}, R., {De Pr{\'a}}, M.~N., {et~al.} 2025, Nature Astronomy, 9, 230, \dodoi{10.1038/s41550-024-02433-2}

\bibitem[{{Rein} \& {Liu}(2012)}]{2012A&A...537A.128R}
{Rein}, H., \& {Liu}, S.~F. 2012, \aap, 537, A128, \dodoi{10.1051/0004-6361/201118085}

\bibitem[{{Rein} \& {Spiegel}(2015)}]{2015MNRAS.446.1424R}
{Rein}, H., \& {Spiegel}, D.~S. 2015, \mnras, 446, 1424, \dodoi{10.1093/mnras/stu2164}

\bibitem[{{Rickman} {et~al.}(2014){Rickman}, {Wi{\'s}niowski}, {Wajer}, {Gabryszewski}, \& {Valsecchi}}]{Rickman2014}
{Rickman}, H., {Wi{\'s}niowski}, T., {Wajer}, P., {Gabryszewski}, R., \& {Valsecchi}, G.~B. 2014, \aap, 569, A47, \dodoi{10.1051/0004-6361/201423966}

\bibitem[{{Schambeau} {et~al.}(2026){Schambeau}, {Kelley}, {Womack}, {Lilly}, {Kareta}, {Faggi}, {Harrington Pinto}, {Micheli}, {Bockel{\'e}e-Morvan}, {Fern{\'a}ndez}, {McKay}, {Pinilla-Alonso}, {Licandro}, {Beck}, {Villanueva}, {Bauer}, {Feaga}, {DiSanti}, \& {Wierzchos}}]{2026PSJ.....7..137S}
{Schambeau}, C.~A., {Kelley}, M. S.~P., {Womack}, M., {et~al.} 2026, \psj, 7, 137, \dodoi{10.3847/PSJ/ae685c}

\bibitem[{{Snodgrass} {et~al.}(2025){Snodgrass}, {Holt}, {Kelley}, {Opitom}, {Guilbert-Lepoutre}, {Knight}, {Kokotanekova}, {Jehin}, {Mazzotta Epifani}, {Migliorini}, {Tubiana}, {Micheli}, \& {Farnocchia}}]{2025MNRAS.541L...8S}
{Snodgrass}, C., {Holt}, C.~E., {Kelley}, M. S.~P., {et~al.} 2025, \mnras, 541, L8, \dodoi{10.1093/mnrasl/slaf046}

\bibitem[{{Trigo-Rodr{\'\i}guez} {et~al.}(2008){Trigo-Rodr{\'\i}guez}, {Garc{\'\i}a-Melendo}, {Davidsson}, {S{\'a}nchez}, {Rodr{\'\i}guez}, {Lacruz}, {de Los Reyes}, \& {Pastor}}]{2008A&A...485..599T}
{Trigo-Rodr{\'\i}guez}, J.~M., {Garc{\'\i}a-Melendo}, E., {Davidsson}, B.~J.~R., {et~al.} 2008, \aap, 485, 599, \dodoi{10.1051/0004-6361:20078666}

\bibitem[{{Vincent}(2014)}]{2014acm..conf..565V}
{Vincent}, J. 2014, in Asteroids, Comets, Meteors 2014, 565

\bibitem[{{Volk} \& {Malhotra}(2008)}]{2008ApJ...687..714V}
{Volk}, K., \& {Malhotra}, R. 2008, \apj, 687, 714, \dodoi{10.1086/591839}

\bibitem[{{Wagner} \& {Schleicher}(1997)}]{1997Sci...275.1918W}
{Wagner}, R.~M., \& {Schleicher}, D.~G. 1997, Science, 275, 1918, \dodoi{10.1126/science.275.5308.1918}

\bibitem[{{Wetherill}(1967)}]{Wetherill1967}
{Wetherill}, G.~W. 1967, \jgr, 72, 2429, \dodoi{10.1029/JZ072i009p02429}

\bibitem[{{Wetherill} \& {Cox}(1985)}]{1985Icar...63..290W}
{Wetherill}, G.~W., \& {Cox}, L.~P. 1985, \icarus, 63, 290, \dodoi{10.1016/0019-1035(85)90012-0}

\end{thebibliography}

\bibliographystyle{aasjournal}

\end{document}